\documentclass{elsart3p}
\usepackage{times,mathptmx}
\usepackage[numbers,sort&compress]{natbib}
\usepackage{amssymb}
\usepackage[final]{graphicx}
\usepackage{units}
\usepackage{amsmath}
\def\lsi{\raise0.3ex\hbox{$<$\kern-0.75em\raise-1.1ex\hbox{$\sim$}}}
\def\gsi{\raise0.3ex\hbox{$>$\kern-0.75em\raise-1.1ex\hbox{$\sim$}}}

\def\be{\begin{equation}}
\def\ee{\end{equation}}
\def\ba{\begin{eqnarray}}
\def\ea{\end{eqnarray}}

\begin{document}
%%%%%%%%%%%%%%%%%%%%%%%%%%%%%%%%%%%%%%%%%%%%%%%%%%%%%%%%%%%%%%%%%%%%%%

\begin{frontmatter}
\title{Quantum scale invariance on the lattice}

\author[CH]{Mikhail Shaposhnikov}
\ead{Mikhail.Shaposhnikov@epfl.ch} and
\author[RU]{Igor Tkachev}
\ead{tkachev@ms2.inr.ac.ru}

\address[CH]{
  Institut de Th\'eorie des Ph\'enom\`enes Physiques,
  \'Ecole Polytechnique F\'ed\'erale de Lausanne,
  CH-1015 Lausanne, Switzerland}

\address[RU]{Institute for Nuclear Research of the Russian Academy of Sciences,
60th October Anniversary prospect, 7a, 117312, Moscow, Russia}
\date{\today}

\begin{abstract}
We propose a scheme leading to a non-perturbative definition of
lattice field theories which are scale-invariant on the quantum level.
A key idea of the construction is the replacement of the lattice
spacing by a propagating dynamical field -- the dilaton. We describe
how to select non-perturbatively the phenomenologically viable
theories where the scale invariance is broken spontaneously. Relation
to gravity is also discussed.
\end{abstract}


\end{frontmatter}

%%%%%%%%%%%%%%%%%%%%%%%%%%%%%%%%%%%%%%%%%%%%%%%%%%%%%%%%%%%%%%%%%%%%%%

\section{Introduction}
\label{sec:intro}
It is believed that the scale invariance, existing in {\em classical}
field theories without dimensionfull parameters, is generically broken
in quantum field theory  (for a review see \cite{coleman}). The known
exceptions include finite theories such as $N=4$ super Yang-Mills
\cite{Sohnius:1981sn} or other classes of supersymmetric theories
\cite{Howe:1983wj}, which, however, are lacking an immediate
phenomenological relevance.  The standard regularisations of the
ultraviolet infinities introduce one type of mass scale or another and
this scale makes its way into the renormalized theory. The  explicit 
breaking of scale invariance (SI) or its anomalous breaking by quantum
effects leads to severe fine-tuning problems, facing the realistic
models of particle physics. One of them is the problem of stability of
the weak scale against quadratic quantum corrections and the second
one is the cosmological constant problem, associated with quartic
divergences. 

A quantum field theory (gravity included) with {\em exact}, but
spontaneously broken SI would solve the above-mentioned problems. The
scale-invariance, existing on quantum level, forbids  quadratic or
quartic divergences, exactly in a way the gauge invariance keeps the
photon massless in the Quantum Electrodynamics. The spontaneous
breaking of the dilatational symmetry would introduce all different
mass scales, observed in Nature (including the mass of the Higgs
boson, the QCD scale $\Lambda$ and alike, the Newtons gravity
constant, etc.) through the vacuum expectation value of the dilaton
field. Moreover, a combination of the ideas of SI with those of
unimodular gravity  \cite{vanderBij:1981ym,Wilczek:1983as,Zee:1983jg}
leads to a possible explanation of primordial inflation and of late
acceleration of the universe \cite{Shaposhnikov:2008xb}. 

In \cite{Shaposhnikov:2008xi} a perturbative way for construction of a
new class of quantum theories, where SI is exact, but broken
spontaneously, was suggested. The similar procedure for keeping the
local conformal symmetry intact at the quantum level was proposed
earlier in \cite{Englert:1976ep}.  The basic idea is as follows. To
have a quantum theory which is scale invariant the renormalisation
procedure must not introduce any dimensionfull parameter. This may be
achieved, in  dimensional regularisation of \cite{'tHooft:1972fi}, by 
replacing the 't~Hooft-Veltman normalization scale $\mu$ by an
appropriate combination of dynamical quantum fields in such a way
that the SI is preserved in any space-time dimension $d=4-2\epsilon$. 
Clearly, this introduces new interactions when $\epsilon \neq 0$.
Though their strength is suppressed by $\epsilon$, they leave a trace
in the renormalized theory, as their combination with the poles in
$\epsilon$ coming from counter-terms leads to finite contributions.
The  procedure described above is only possible if the SI is
spontaneously broken (otherwise the perturbative expansion is ill
defined), but this is required anyway by phenomenological
considerations.

If the construction of quantum scale-invariant theories based on
dimensional regularisation is self-consistent, then the result should
have a more general character and other renormalisation schemes,
leading to quantum dilatational invariance, should exist.  Lattice
regularisation plays a special role in construction and studies of
quantum field theories since it allows for non-perturbative approach.
The aim of the present letter is to attempt a non-perturbative lattice
construction of the quantum SI theories.

The paper is organized as follows. In Section \ref{sec:tmodel} we
present the idea of lattice regularisation of SI theories with the use
of  an example of a simple scalar theory. In Section \ref{sec:gravity}
we will discuss the inclusion of gravity. Section \ref{sec:concl} is
conclusions.

\section{Lattice spacing as a dynamical field}
\label{sec:tmodel}
To present our main idea, we start with a simple scalar field model,
given by the action
\begin{eqnarray}
\label{S} 
&&{S}= \int d^4x \; {\cal L},\\ 
&&{\cal L}=\frac{1}{2}\left[(\partial_\mu\chi)^2 +(\partial_\mu h)^2
\right]-V(h,\chi),\\
 \label{L}
&&V(h,\chi)= \lambda\left(h^2-\zeta^2\chi^2\right)^2 +\beta\chi^4~.
 \label{V}
\end{eqnarray}
The fields  $h$ and $\chi$ can be thought of as the Higgs field and
the dilaton correspondingly. Here $\lambda$ and $\zeta$ are
dimensionless coupling constants. The theory is scale-invariant at the
classical level: the action does not change under substitution $h(x)
\to \Omega h(\Omega x)$ and $\chi(x) \to \Omega \chi(\Omega x)$, where
$\Omega$ is a parameter of dilatation transformation. If $\beta > 0$,
the vacuum state of the theory $h=\chi=0$ is scale-invariant. If
$\beta < 0$, the theory does not have a ground state.  The vacuum
state with spontaneously broken SI appears if $\beta=0$ and
corresponds to some point in the flat direction $h^2-\zeta^2\chi^2=0$.
The specific values of coupling constants and of the vacuum
expectation value of the dilaton $\chi_0$ are not important for what
follows.\footnote{The choice required  by  phenomenological considerations
corresponds to   $\chi_0 \simeq M_{\rm P}$, where $M_{\rm P}$ is the
Planck mass, and $\zeta \sim v/M_{\rm P}\sim 10^{-16} \lll 1$, where
$v$ is the electroweak scale.}

The scale-invariant perturbative renormalisation of this model has
been considered in \cite{Shaposhnikov:2008xi}. It was shown there 
that if the  parameter $\mu$ of dimensional regularisation is replaced
by an appropriate combination of dynamical fields, then the
counter-terms can be chosen in such a way that the divergences are
removed and the classical flat direction, existing at $\beta=0$, is
not lifted by quantum corrections. A choice, motivated by 
cosmological considerations \cite{Shaposhnikov:2008xb}, reads
\be
\mu^{2\epsilon} \to \left[\omega^2\right]^{\frac{\epsilon}{1-\epsilon}}~,
\label{GR}
\ee
where $\omega^2 \equiv \left(\xi_\chi \chi^2 + \xi_h h^2\right)$ and 
$\xi_\chi,~\xi_h$ are the couplings of the scalar fields to the Ricci
scalar, see section \ref{sec:gravity}. The low energy theory contains
a massive Higgs field and the massless dilaton, the latter being a Goldstone
boson of the broken scale invariance.\footnote{This particle has only 
derivative couplings to matter and thus escapes experimental bounds, 
given the small value of $\zeta$.} A more familiar example of a
scale-invariant quantum theory with spontaneous breaking of SI, which
appears in standard renormalisation procedure, is $N=4$ supersymmetric
Yang-Mills theory \cite{Sohnius:1981sn}.

The requirement that the scale invariance must be spontaneously broken
is essential for perturbative construction. Only in this case the
perturbative computations can be done.  Moreover, if both the ground
state and the action respect SI, the resulting theory does not contain
any mass scale and thus cannot be accepted phenomenologically.

Let us turn now to lattice regularisation (LR). Consider some
space-time lattice of points, with oriented edges connecting them. The
space-time is taken to be infinite at the moment. Finite difference
scheme can be constructed in many different ways leading to the same
result in continuum limit. To simplify the subsequent discussion let
us denote by $\phi_{i+}$ and $\phi_{i-}$ the values of a generic field
$\phi$ at space-time points corresponding to the i-th edge end.  
The field difference and the field strength associated to
this edge is $\Delta \phi_i = \phi_{i+} - \phi_{i-}$ and $\phi_i =
(\phi_{i+} + \phi_{i-})/2$ correspondingly. We assume that the length
of the lattice edges $a_i$  may vary in space and time. Then the
classical lattice action of the model eq. (\ref{L})  becomes
\be
S =  \sum_{ i} a_i^4\; \left[\frac{\Delta \chi_{ i}^2 + 
\Delta h_{ i}^2}{2a_i^2} 
- V(h_{ i},\chi_{ i})\right]~.
\label{Lat1}
\ee

Clearly, this regularisation breaks explicitly the scale invariance.
It introduces the lattice spacing $a_i$, playing the role of the
inverse ultraviolet cutoff in quantum theory. Therefore, the only
possibility to construct a scale invariant theory at quantum level in
the framework of LR, is to promote the lattice spacing to a dynamical
field $\phi$, 
\be
a_i^{-1} = \eta \phi_i~,
\label{dynlatt}
\ee
where $\eta$ is some constant, determining how fine is the grid. A
possibility to have an ultraviolet cutoff as a dynamical field in SI
theories was discussed previously  in \cite{Wetterich:1987fm}.

The $\phi$ could be a new field, not present yet in eq. (\ref{L}). 
However, we can assume that our starting action contains already all
scalar fields of the theory. Then, in general, $\phi$ is a function of
$h$ and $\chi$. We take it to be $\phi^2=\omega^2$, in analogy with
the previous discussion, see eq. (\ref{GR}). Then the action on the 
lattice becomes
\be
S =  \sum_{ i}\; \left[\frac{\Delta \chi_{i}^2 + 
\Delta h_{i}^2}{2\omega_{ i}^2\eta^2}
- \frac{V(h_{i},\chi_{ i})}{\omega_{i}^4\eta^4}\right].
\label{ConstSp}
\ee
It is invariant under the scale transformations 
\be
 h_i \to \Omega h_i,~~~\chi_i \to \Omega \chi_i~.
\ee

The quantum SI system can be defined by the Euclidean partition
function,
\be
Z = \prod_i \int \frac{d\chi_i dh_i}{\sigma_i^2}\;  e^{-S_E},
\label{measure}
\ee
where we introduced the scale-invariant path integral measure and made
the redefinition of the fields (in such a way that $\omega_i^2\to
\sigma_{i}^2=h_{i}^2 + \chi_{i}^2$) and coupling constants accounting
for renormalisation as follows:
\be
S_E = 
\sum_{ i}\; \left[\frac{A\Delta \chi_{i}^2 + 
B \Delta h_{i}^2}{2\sigma_{ i}^2}
+\frac{C \chi_{i}^4 + D \chi_{i}^2 h_{i}^2 + E
h_{i}^4}{\sigma_{i}^4}\right]~,
\label{latt}
\ee
where $A,~B,~C,~D$ and $E$ are $5$ arbitrary parameters.  If the
theory is considered in a finite space-time volume, the summation in
(\ref{latt}) is limited by the total number $N$ of the lattice edges.

The naive continuum limit of the lattice theory (the physical volume
is fixed) is achieved by the scaling $\eta \to \kappa \eta$, $(A,B)\to
(A,B)/\kappa^2$,  $(C,D,E)\to (C,D,E)/\kappa^4$, $N\to\kappa^4 N$,
$\kappa\to\infty$, corresponding to a finer covering of the space by
the lattice points. 

For numerical lattice simulations another choice of variables can be
more convenient. For example, the change of variables $\chi =
\exp(\theta) \cos(\phi)$, $h = \exp(\theta) \sin(\phi)$ will simplify
the kinetic and potential terms.

The phenomenologically interesting theories are those where the
scale invariance is spontaneously broken. This does not necessarily
happen for all possible choices of parameters in the lattice action
(\ref{latt}). To select a class of theories with SI spontaneously
broken, one can construct an effective potential
$V_{eff}(\chi,h)$, given by
\ba
\label{Veff}
&&\exp\left(-N \sigma^{-4} V_{eff}(\chi,h)\right) =\\
\nonumber
&&\frac{1}{Z} \prod_i 
\int \frac{d\chi_i dh_i}{\sigma_i^2}  
\delta\left(1 - \frac{\bar\chi}{\chi}\right) 
\delta\left(1 - \frac{\bar h}{h}\right)\;  e^{-S_E}~,
\ea
where $\sigma^2 = \chi^2 + h^2$ and
\be
\bar\chi = \frac{1}{N} \sum_{i}\chi_i,~~~\bar h = \frac{1}{N}
\sum_{i}h_i~.
\ee
Note that the factor of $N\sigma^{-4}$ appears in the l.h.s. of eq.
(\ref{Veff}) because we have to recover integration over classical
space-time. Due to SI, the effective potential is  $V_{eff}(\chi,h) =
\sigma^{4} W(x)$, where $x = h/\chi$. To get a spontaneous breaking of
SI the parameters of the action must be chosen in such a way that the
minimum of $W(x)$ is achieved at some point $x_0$ such that
$\sigma\neq 0$ and  
\be
\lim_{N\to\infty}W(x_0)=0~.
\label{surf}
\ee
In terms of field variables the minimum of $W(x)$ correspond to the
flat direction, $h= x_0 \chi$. This condition singles out some $4$ -
dimensional surface in $5$-dimensional parameter-space, as it happens
in the classical theory or in quantum theory with perturbative SI
renormalisation prescription. 

Several important comments are now in order.

(i) A continuum limit of the lattice model (\ref{latt}) may lead to a
non-interacting trivial field theory (as in the case of 
$\lambda\phi^4$ theory  \cite{Wilson:1973jj,Frohlich:1982tw,
Luscher:1987ay}) even on the surface corresponding to spontaneous
breaking of scale-invariance. In this case the action (\ref{latt})
cannot be considered as fundamental and would rather correspond to an
effective field theory, valid below some energy scale. This effective
theory is exactly scale invariant, with SI spontaneously broken. An
extreme point of view which does not require an existence of continuum
limit is that the space-time is in fact discrete. 

(ii) If the parameters of the lattice action are such that eq.
(\ref{surf}) does not hold, the ground state of the theory is
scale-invariant. The lattice simulations in these case can be used to
determine at the non-perturbative level the anomalous dimensions of
different composite operators. 

(iii) The measure for functional integral cannot be fixed on the
grounds of scale-invariance only.  The eq. (\ref{measure}) provides
just one of the possible examples. The freedom in the choice of the
measure and in the expression of the field $\phi$ through $h$ and
$\chi$ corresponds to the freedom of renormalisation prescription in
perturbative SI, discussed in \cite{Shaposhnikov:2008xi}. 

(iv) One can add to the lattice action (\ref{Lat1}) infinite number
of terms, containing higher dimensional operators suppressed by the
lattice spacing, for example  $\sum_{ i}\; a_i^6 \chi_i^6$. These 
will be transformed to $F \sum_{ i}\; \chi_i^6/\sigma_i^6$ by our
procedure. We expect these terms be irrelevant in the continuum limit,
since $F$ scales as $F/\kappa^6$ whereas the number of space-time
points grows as $\kappa^4$ only.

(v) By introducing some artificial constant length scale $a_0$,  we
can formally rewrite (\ref{ConstSp}) as an action on the lattice with
a constant spacing. Then in the limit  $a_0\to 0$ the continuum
Lagrangian corresponding to the action eq. (\ref{ConstSp}) is 
%(though we wrote the Planck scale below, it could be an arbitrary mass 
%parameter)
\be
{\cal L}_0 = \frac{1}{2}\frac{ M_{\rm P}^2}{\omega^2}
\left[(\partial_\mu\chi)^2 +(\partial_\mu h)^2\right] 
-\frac{M_{\rm P}^4}{\omega^4}\; V(h,\chi)~,
\label{ConstSpCont}
\ee
where $M_{\rm P}^2$ should be understood as the vacuum expectation
value of $\omega$. The Lagrangian (\ref{ConstSpCont}) corresponds to a
highly non-linear and non-renormalisable theory. It is intriguing that
it looks like the field theory part of the scale-invariant Lagrangian
of the scalar fields coupled to gravity after transition to the
Einstein frame \cite{Shaposhnikov:2008xb}. One could have started with
the bare action eq. (\ref{ConstSpCont}) right away.  But then the
intriguing relation of this theory to a dynamical discrete space-time
would have been lost.

(vi) Clearly, the notion of the classical space and time can appear in
our approach only if the scale-invariance is spontaneously broken and
the field $\sigma$ acquires a non-zero expectation value.  The arising
space-time is flat in the case of a ground state with $\sigma = $
const.  One can imagine several possibilities for emerging of gravity.
First of all, an effective low energy theory derived from
eqs.~(\ref{measure})-(\ref{latt}) may already contain the Einstein or
unimodular gravity. Put it in other words, the correlator of some
composite spin 2 operators, constructed from the fields $h$ and
$\chi$, may have a massless pole corresponding to the graviton. If
true, the Hilbert-Einstein action is effective rather than
fundamental.  In alternative situation one may have to add to the
action (\ref{latt}) an extra term describing the gravity in a
conventional way, as discussed below.

%%%%%%%%%%%%%%%%%%%%%%%%%%%%%%%%%%%%%%%%%%%%%%%%%%%%%%%%%%%%%%%%%%%%%%
\section{Inclusion of scale-invariant gravity} 
\label{sec:gravity}
The lattice spacing, represented by a quantum field, is nothing but a
form of space-time quantization, which could lead eventually to
quantum gravity. This approach to the lattice formulation of gravity
goes back to Regge \cite{Regge:1961px} (for reviews see, e.g. 
\cite{David:1992jw,Ambjorn:1996ny,Regge:2000wu}).  The basic idea is
that a curved space can be approximated by a collection of piecewise
flat manifolds or simplices (triangles for the case of 2-dimensional
surfaces, tetrahedra and pentatopes for the case of 3 and 4-dimensions
respectively). A $d$-dimensional simplex has $d-1$-dimensional
``sides'' or ``faces''. In turn, faces are bounded by a
$d-2$-dimensional ``hinges''. (In the triangulation of a 2-dimensional
surface a hinge is a point.) In a piecewise flat manifold the
curvature is concentrated on hinges \cite{Regge:1961px}, so they play
special role in the formalism.  This approach avoids the use of
coordinates. In it the lengths of 1-dimensional edges of simplices
play the role of dynamical variables, exactly what we need for
construction of scale-invariant quantum theories. So, we identify the
length of every edge of 4d simplex with inverse of our field $\sigma$
as in (\ref{dynlatt}).

The standard Einstein action in Regge formalism is written as
\be
M_{\rm P}^2 \int d^4x \sqrt{g}R ~~\rightarrow~~ 
M_{\rm P}^2 \sum_{hinges~i} V_i \delta_i , 
\label{regge}
\ee
where $V_i$ is the area of two dimensional hinge and $\delta_i$ is the
deficit angle there, equal to $2\pi$ minus the sum of the dihedral
angles between the faces of the simplices meeting at that hinge.

Of course, this expression is not scale-invariant due to the presence
of explicit Planck mass. To get the SI action, we replace $M_{\rm P}$
by a dynamical scalar field which can be chosen to be $G\sigma$, where $G$
is some constant. We arrive to the Brans-Dicke like action
\be
S_G =  G \sum_{hinges~i} \delta_i V_i \sigma_i^2~.
\label{reggeSI}
\ee
Under the global scale transformations, the deficit angle does not
change, while the area scales as $V_i \propto \sigma_i^{-2}$.
Continuum limit (volume is fixed) corresponds to $G \to G/\kappa^2$,
$\kappa\to\infty$ (number of hinges scales as $\kappa^2$).

The sum in eq. (\ref{reggeSI}) can be explicitly evaluated in the
Dynamical Triangulation (DT) approach
\cite{Kazakov:1985ds,David:1984tx,Ambjorn:1985az} which is a variant
of the Regge calculus where all simplexes are equilateral.  Also, the
separation of metric variables and the scalar field is not clear if
one sticks to the original Regge version, but it becomes transparent
if one changes to DT where all the geometric degrees of freedom are in
the connectivity of the triangulation.  Since in an equilateral
simplex in d-dimensional space the angle between two faces sharing a
common hinge is given by $\arccos(1/d)$, the deficit angle is given by

\be
\delta_i = 2\pi - n_i\, \arccos(1/d)~,
\ee
where $n_i$ denotes the number of simplices incident on an $i$-th hinge. 
In $d$ dimensions a hinge has dimensionality $j=d-2$ and since hinges are also 
equilateral, their volume is given by
\be
V_i = \frac{a^j \sqrt{j+1}}{j!\sqrt{2^j}}~,
\ee
which in four dimensions becomes $V_i = \sqrt{3}a^2/4$.  By replacing
$a_i^{-1}\to\sigma_i$ we obtain for the gravitational action
\ba
\nonumber
S_R &&= G \sum_{hinges~i} V_i \delta_i \sigma^2_i\\
&&= G \sum_{hinges~i}\frac{\sqrt{3}}{4} \left(2\pi -  
n_i \arccos\left(\frac{1}{4}\right)\right) ~,
\label{DTA}
\ea
which reduces to $S_R = k_2 N_2 - k_4N_4$, where $N_2$ and $N_4$ are
the numbers of hinges and simplexes in a given simplical manifold,
while $k_2$ and $k_4$ are some bare constants, cf.
\cite{Ambjorn:1991pq,Ambjorn:2008wc}.

Other matter fields can be added to the Regge formalism as well (see,
e.g. \cite{David:1992jw,Ambjorn:1996ny,Regge:2000wu}) and the theory 
can be made scale-invariant
by the replacement (\ref{dynlatt}).  As the lengths of the of edges
are invariant under general coordinate transformations, the invariant
measure can be chosen as in eq.  (\ref{measure}) with the appropriate
powers of $\sigma_i$.

%%%%%%%%%%%%%%%%%%%%%%%%%%%%%%%%%%%%%%%%%%%%%%%%%%%%%%%%%%%%%%%%%%%%%%
\section{Conclusions}
\label{sec:concl}

In this letter we argued that it is possible to identify an
ultraviolet cutoff, needed for regularisation of particle physics
models, with a dynamical field -- the dilaton. A concrete realisation
of this suggestion with the use of lattice field theories is proposed.
This construction may lead to existence of a new class of theories,
which are perturbatively and non-perturbatively scale invariant at the
quantum level. In this type of theories the problem of stability of
the Higgs mass hierarchy and of the cosmological constant are solved
automatically.

An intriguing feature of this proposal is that it indicates that a
solution of two outstanding problems, namely the origin of different
mass scales in particle physics (spontaneous breaking of quantum scale
invariance) and quantization of space-time (dynamical cutoff), may be
intricately related.

Clearly, a lot of work is required to see if this suggestion goes
through. The important question is whether the non-perturbative
continuum limit of spontaneously broken SI lattice models, described
in this paper, leads to non-trivial theories. The lattice simulations
with well developed tools and techniques would allow to approach this
question.

%%%%%%%%%%%%%%%%%%%%%%%%%%%%%%%%%%%%%%%%%%%%%%%%%%%%%%%%%%%%%%%%%%%%%%

\section*{Acknowledgements} 
This work was supported by the Swiss National Science Foundation. We
thank J. Ambjorn, S. Khlebnikov, M. Laine, H. Meyer, A. Rosly, 
K. Rummukainen and S. Sibiryakov
for valuable comments. I.T. thanks EPFL, where this work was done, for
hospitality.

%\bibliography{all,bookrefs}
%\bibliographystyle{h-elsevier3-s}

\end{document}